\documentclass[11pt, a4paper]{article}
\pdfoutput=1

\usepackage[height=8.85in,width=6.75in]{geometry}

\usepackage{chngcntr}
\counterwithin{equation}{section}

\usepackage{graphicx,rotating}     
\usepackage[bookmarksopen,colorlinks=true,linkcolor=light_blue,
citecolor=light_pink,urlcolor=dark_red,linktoc=all]{hyperref}

\usepackage{amsmath}
\usepackage{mathtools}
\usepackage{amsfonts}
\usepackage{amssymb}
\usepackage{slashed}
\usepackage{braket}
\usepackage{bm}
\usepackage{bbm}
\usepackage{cite}
\usepackage{comment}
\usepackage{mathrsfs}
\usepackage{xcolor}
\usepackage[utf8]{inputenc}
\usepackage[footnotesize]{caption}
\usepackage{bbold}
\usepackage{xfrac}
\usepackage{physics}
\usepackage{extarrows}

\usepackage[lf]{Baskervaldx} 
\usepackage[bigdelims,vvarbb]{newtxmath} 
\usepackage[cal=boondoxo]{mathalfa} 

\newcommand{\Mpl}{M_{\text{Pl}}}

\def\thefootnote{\fnsymbol{footnote}}
\usepackage{multicol}
\definecolor{dark_red}{rgb}{0.7, 0., 0.}
\definecolor{light_pink}{rgb}{1,0.4,0.4}
\definecolor{light_blue}{rgb}{0.284602,0.317763,0.963947}
\definecolor{forestgreen}{HTML}{228B22}
\definecolor{ochre}{HTML}{CCAA2B}

\begin{document}
\hypersetup{pageanchor=false}
\begin{titlepage}

    \begin{center}

        \hfill KEK-TH-2639\\
        \hfill TU-1238\\
        \hfill CTPU-PTC-24-22\\

        \vskip .7in
            {\Huge \bfseries Thermalization and hotspot formation\\
                around small primordial black holes\\}
        \vskip .6in
            {\Large Minxi He$^a$, Kazunori Kohri$^{b,c,d,e}$, Kyohei Mukaida$^{c}$, Masaki Yamada$^{f,g}$}

        \vskip .3in
        \begin{tabular}{ll}
            $^a$ & \!\!\!\!\!\emph{Particle Theory and Cosmology Group, Center for Theoretical Physics of the Universe, } \\[-.3em]
                 & \!\!\!\!\!\emph{Institute for Basic Science (IBS),  Daejeon, 34126, Korea}                             \\
            $^b$ & \!\!\!\!\!\emph{Division of Science, National Astronomical Observatory of Japan (NAOJ), and SOKENDAI, }               \\[-.3em]
                 & \!\!\!\!\!\emph{2-21-1 Osawa, Mitaka, Tokyo 181-8588, Japan}                                           \\
            $^c$ & \!\!\!\!\!\emph{Theory Center, IPNS, KEK, 1-1 Oho, Tsukuba, Ibaraki 305-0801, Japan}                   \\
            $^d$ & \!\!\!\!\!\emph{QUP (WPI), KEK, 1-1 Oho, Tsukuba, Ibaraki 305-0801, Japan}                            \\
            $^e$ & \!\!\!\!\!\emph{Kavli IPMU (WPI), University of Tokyo,
            Kashiwa, Chiba 277-8568, Japan}                       \\
            $^f$ & \!\!\!\!\!\emph{Frontier Research Institute for Interdisciplinary Sciences, Tohoku University,}               \\[-.3em]
                 & \!\!\!\!\!\emph{ 6-3 Azaaoba, Aramaki, Aoba-ku, Sendai 980-8578, Japan}                               \\
            $^g$ & \!\!\!\!\!\emph{Department of Physics, Tohoku University, 6-3 Azaaoba, Aramaki, Aoba-ku, Sendai 980-8578, Japan }              \\
        \end{tabular}

    \end{center}

    \vskip .5in

    \begin{abstract}
        \noindent
We quantitatively analyze a basic question: what is the stationary solution of the background plasma temperature profile around a black hole (BH)? One may naively expect that the temperature profile continuously decreases from the Hawking temperature at the surface of the BH towards an outer region. We show analytically and numerically that this is not the case because local thermal equilibrium cannot be maintained near the surface of the BH and also because the high-energy particles emitted from Hawking radiation cannot be instantaneously thermalized into the background plasma. The temperature profile has a plateau within a finite distance from the BH, and even the overall amplitude of background temperature at a distance far away from the BH is significantly suppressed compared with the naive expectation. The main reason for these counterintuitive results comes from the fact that the size of the BH is too small that particles of Hawking radiation goes far away within the typical time scale of interactions. 
    \end{abstract}

\end{titlepage}

\tableofcontents
\thispagestyle{empty}
\renewcommand{\thepage}{\arabic{page}}
\renewcommand{\thefootnote}{$\natural$\arabic{footnote}}

\setcounter{footnote}{0}
\newpage
\hypersetup{pageanchor=true}

\section{Introduction}
\label{sec:introduction}

A black hole (BH) evaporates via the Hawking radiation~\cite{Hawking:1974rv,Hawking:1975vcx} from its surface.
The distribution of radiated particles is characterized by the Hawking temperature.
Its hidden thermodynamics nature is further supported not only by the fact that a BH fulfills the thermodynamic laws~\cite{Bekenstein:1972tm,Bekenstein:1973ur,Bardeen:1973gs} (see also \cite{Carlip:2014pma} for a review and references therein) but by the confirmation of the Bekenstein--Hawking entropy based on statistical mechanical calculation of string theory~\cite{Susskind:1993ws,Susskind:1994sm}.
The thermodynamics of the system including the BH and a thermal plasma in a finite volume implies that the whole system can be in equilibrium at the Hawking temperature, although, in an infinite box, the system is unstable due to the negative specific heat of BH.
History has shown that understanding the thermodynamic property of a BH system provides a non-trivial and important progress for theoretical physics.

In the present paper, we consider a system with a BH put in a thermal plasma whose temperature is much lower than the Hawking temperature and investigate its stationary solution at a late time (but at a time before the BH completely evaporates).
The radiation emitted as a Hawking radiation interacts with the ambient plasma, makes it hot locally around the BH, and the system eventually reaches the stationary solution.
One might expect that the local temperature close to the horizon becomes as large as the Hawking temperature at the BH surface~\cite{Das:2021wei,Hamaide:2023ayu}.

However, one has to note that, even if particles have a thermal distribution, it does not mean that they are in thermal equilibrium.
For example, the cosmic microwave background (CMB) is observed as a blackbody-radiation spectrum at present, though it is decoupled from the thermal plasma at the last scattering surface.
In fact, in the present paper, we clarify that the thermal equilibrium can not be maintained near the surface of the BH and the temperature of the background plasma in the vicinity of the surface of the BH cannot coincide with the Hawking temperature.
This is implicitly included in our previous quantitative analysis in Ref.~\cite{He:2022wwy}.

\begin{figure}[ht]
    \centering
    \includegraphics[width=0.9\linewidth]{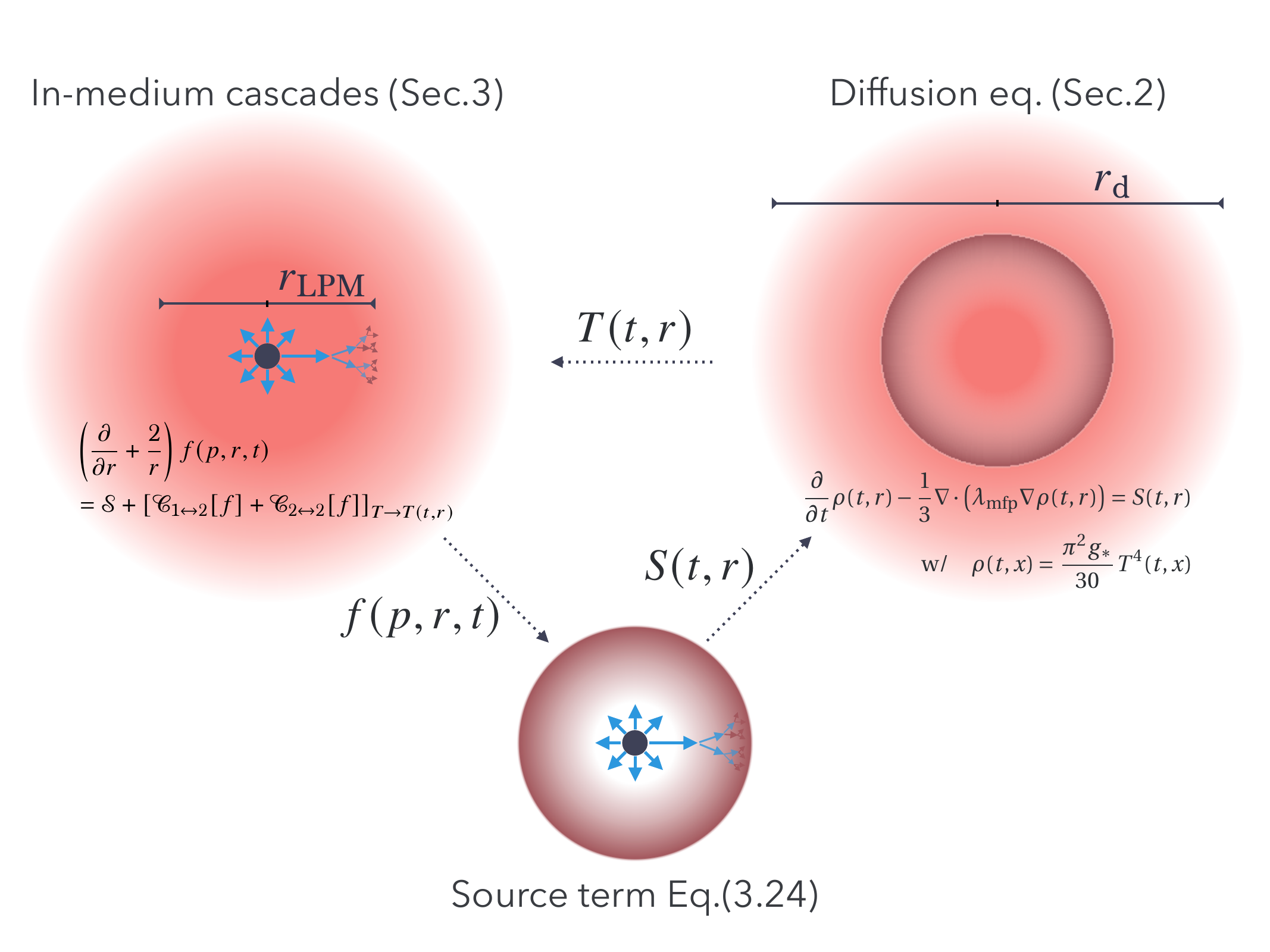}
    \caption{
        Schematic figure of how we formulate the hotspot formation by the evaporating small PBH.
        \textit{Top right:} in the presence of source term, $S(t,r)$, given in the brown shade, the temperature profile, $T(t,r)$, given in the red shade, is obtained by solving the diffusion equation.
        The details of this macroscopic description based on local thermal equilibrium are given in Sec.~\ref{sec:radtr}.
        \textit{Top left:} as we will see shortly, the local thermal equilibrium cannot be attained near BH because of its extreme compactness.
        Hence, we need to solve the kinetic equation.
        The bottleneck for the thermalization of Hawking radiation, given in blue solid arrows, is in-medium cascades, known to be suppressed by the LPM effect.
        For a given $T(t,r)$, one may obtain the phase space density, $f(p,r,t)$, by (numerically) solving the Boltzmann equation.
        The details of effective Boltzmann equation including the LPM effect are discussed in Sec.~\ref{sec:thermalization}.
        \textit{Bottom:} for a given $f(p,r,t)$, one may readily compute the fraction of the thermalized population, which gives the source term $S(t,r)$. 
        This will be then plugged into the diffusion equation of locally thermal plasma (top right), which affects $T(t,r)$. In this way, we need to take all these processes self-consistently.
        \textit{Note:} as we discuss in Sec.~\ref{subsec:num}, we do not have to solve the Boltzmann equation repeatedly for each $T(t,r)$ because the time and spatial dependence of $T(t,r)$ can be factored out.
    }
    \label{fig:schematic}
\end{figure}

To be more specific, we clarify the following non-trivial properties of finite-temperature system around a small BH put in a thermal plasma whose temperature is much lower than the Hawking temperature:
\begin{itemize}
\item
mean free path in the ambient plasma near the BH is always much longer than the BH radius, and hence the local thermal equilibrium cannot be maintained near the surface of the BH.
\item
high-energy particles emitted as Hawking radiation can be absorbed into the ambient plasma
far away from the BH
\end{itemize}
We discuss these conditions based on the basic theory of radiative transfer and Boltzmann equation. 
The latter includes the effect of interference effect for thermal scattering of highly energetic particles in a relatively low-temperature plasma, known as the Landau-Pomeranchuk-Migdal (LPM) effect~\cite{Landau:1953um,Migdal:1956tc,Gyulassy:1993hr,Arnold:2001ba,Arnold:2001ms,Arnold:2002ja,Besak:2010fb,Kurkela:2011ti}.
We numerically solve the Boltzmann equation and diffusion equation for particles emitted as Hawking radiation (see Fig.~\ref{fig:schematic} for a schematic figure representing the whole processes).
A temperature profile around a BH is given as a stationary solution of the diffusion equation.
The result is qualitatively consistent with our previous work~\cite{He:2022wwy}, but the present paper clarifies the reason why we do not have a local thermal plasma with temperature as high as the Hawking temperature.

We may have to note that the original paper~\cite{Das:2021wei} has mainly focused on the late-time distribution of temperature that is deviated from the $r^{-1/3}$ power law behavior because of the fast evaporation of BH at the last stage of its life.
This is also non-trivial and is interesting feature of temperature profile around an evaporating BH.
For simplicity, we do not revisit this point in the present paper though
its time dependence can be easily included in our analysis.
Throughout this paper,
we focus on the time scale in which we can neglect the time evolution of the BH mass to simplify the analysis. 
Even in this regime, the results drastically change from the naive expectation.

The rest of the paper is organized as follows.
In Sec.~\ref{sec:radtr},
we explain the equation governing radiative transfer and clarify the condition of local thermal equilibrium.
The gradient of the temperature should be smaller than the inverse of the typical length scale determined by the mean free path of particles in the thermal plasma.
Under the assumption of the local thermal equilibrium, one can write down the radiative transfer equations that result in the diffusion equation in a reasonable limit. However, the diffusion equation does not say anything about the boundary condition, which should be imposed consistently with the source term for the background plasma.
In Sec.~\ref{sec:BH}, we discuss those equations for the case with an evaporating BH and explain that it is not in thermal equilibrium even locally near the surface of the BH.
This is because the size of the BH is too small and is much smaller than the mean free path of particles in the thermal plasma.
In Sec.~\ref{sec:thermalization}, we discuss the thermalization of high-energy particles emitted as Hawking radiation.
We first present the qualitative estimate of the thermalization rate of these particles with LPM suppression in Sec.~\ref{subsec:LPM}, and then briefly discuss the temperature profile around an evaporating BH expected from the qualitative discussion. 
In Sec.~\ref{subsec:kineq}, we provide a detailed Boltzmann equation that governs the thermalization of the high-energy particles. 
Once the particles are thermalized into the background plasma, they contribute to the source term for the diffusion equation.
The Boltzmann equation and the diffusion equation are solved by numerical simulations in Sec.~\ref{sec:simulation}. Our results quantitatively clarify how the high-energy particles are thermalized near the BH.
The stationary solution to the diffusion equation is consistent with our qualitative estimation. In particular, the maximal temperature around the BH is significantly lower than the Hawking temperature.
Finally, we conclude our work in Sec.~\ref{sec:discussion}.

\newpage

\newpage
\section{Radiative transfer under local thermal equilibrium}
\label{sec:radtr}

This section summarizes the basics of radiative transfer under local thermal equilibrium.
We first clarify the condition to have well-defined local temperature, \textit{i.e.,} the notion of local thermal equilibrium, in Sec.~\ref{subsec:locth}.
Then, in Sec.~\ref{subsec:radtr}, we derive the governing equation of radiative transfer from the source placed at the origin to its surroundings by assuming local thermal equilibrium.
In Sec.~\ref{subsec:Tprof},
we obtain the temperature profile as an equilibrium solution at a distance far away from external source in a spherically symmetric system.
The boundary condition coming from the external source is discussed in Sec.~\ref{sec:BC}.
Finally, in Sec.~\ref{sec:BH}, we apply the discussion to the case with an evaporating BH that provides a source term for the background diffusion equation from the Hawking radiation.

\subsection{Local thermal equilibrium and mean free path}
\label{subsec:locth}

The notion of local temperature plays the central role throughout this paper since we are interested in a situation where an energetic source, such as an evaporating small BH, heats up the ambient plasma.
This requires some clarification since, at the zero-th order, temperature is well defined in global thermal equilibrium.
Nevertheless, we may still apply the concept of thermodynamics locally as far as the temperature depends on the spatial coordinate $\bm{x}$ so weakly that the particles essentially feel the temperature as if it were constant.

To quantify this consideration, let us take a temperature profile denoted by $T (\bm{x})$.
Consider a box surrounding $\bm{x}_0$ whose size $L$ is small enough to neglect the spacial-dependence of $T(\bm{x}_0)$ but large enough to contain sufficient number of relativistic particles, \textit{i.e.,}
\begin{equation}
    T(\bm{x}_0) \gg \abs{ \bm{\nabla} T (\bm{x}_0) } \times L \,, \qquad
    T^3 (\bm{x}_0) \times L^3 \gg 1 \,.
    \label{eq:box}
\end{equation}
This box of $L^3$ can be regarded as local thermal equilibrium if each particle can interact with each other frequently within the box.
The typical length scale of large angle scatterings is characterized by the mean free path of $\lambda_\text{mfp} \sim 1 / (\alpha^2 T)$ with $g = \sqrt{4 \pi \alpha}$ being a gauge coupling constant for relativistic thermal plasma.
To achieve local thermal equilibrium within the box, we need to guarantee $L \gg \lambda_\text{mfp}$, which, together with the first equation in Eq.~\eqref{eq:box}, leads to the following condition
\begin{equation}
    \alpha^2 \gg \frac{\abs{ \bm{\nabla} T (\bm{x}) }}{ T^2 ( \bm{x} ) } \,.
    \label{eq:cond_locth}
\end{equation}
Here we omit the subscript $0$ in $\bm{x}$ since this condition should be fulfilled anywhere.
Note that the second condition in Eq.~\eqref{eq:box} is satisfied trivially under $L \gg \lambda_\text{mfp}$ when assuming weak coupling $ \alpha \ll 1 $.

Equation \eqref{eq:cond_locth} is the first condition one has to take into account.
Conversely,
the temperature gradient within the length scale of $\lambda_\text{mfp}$ is not meaningful under the local thermal equilibrium.
This changes the picture of thermal profile in the vicinity of a BH, as we will see in Sec.~\ref{sec:BH}.

\subsection{Radiative transfer}
\label{subsec:radtr}

The fundamental building block of the radiative transfer is the specific intensity $I_\nu$, whose definition will be given shortly.
Consider 
an area element $\dd A$ at $\bm{x}$ with its normal being $\hat {\bm n}$.
The amount of energy passes through $\dd A$ within a solid angle $\dd \Omega$ in a direction of $\hat {\bm \Omega}$  
in time $\dd t$ in a frequency interval between $\nu$ to $\nu + \dd \nu$ can be expressed as
\begin{equation}
    \dd E_\nu = I_\nu (t, \bm{x}, \hat{\bm{\Omega}})\,\cos \theta\, \dd A \dd \Omega \dd t \dd \nu \,,
    \qquad
    \cos \theta = \hat {\bm \Omega} \cdot \hat {\bm n}\,.
\end{equation}
The energy flux is readily obtained from
\begin{equation}
    {\bm J} (t, {\bm x}) = \int \dd \Omega \dd \nu \, \hat {\bm \Omega}\, I_\nu ( t, \bm{x}, \hat{\bm{\Omega}} ) \,.
\end{equation}
If we take the direction of ray to coincide with the normal of $\dd A$, \textit{i.e.,} $\hat {\bm \Omega} \cdot \hat {\bm n} = 1$, the energy passes through $\dd A$ per time $\dd t$ defines the corresponding energy density within a box of $\dd t \times \dd A$.
By integrating over the whole direction of ray and frequency, we obtain the energy density
\begin{equation}
    \rho ( t , {\bm x}) = \int \dd \Omega \dd \nu \, I_\nu (t , {\bm x}, \hat {\bm \Omega}) \,.
\end{equation}
One may also express the pressure tensor as
\begin{equation}
    \label{eq:pressure}
    p_{ij} (t, {\bm x}) = \int \dd \Omega \dd \nu\, \hat{\bm \Omega}_i \hat{\bm \Omega}_j \, I_\nu (t, {\bm x}, \hat {\bm \Omega})
    = \delta_{ij} p (t, {\bm x})\,.
\end{equation}
In the second equality, we have used the assumption of local thermal equilibrium, where there is no specific direction.
In this case, the following well-known property immediately follows, \textit{i.e.,} $p = \rho / 3$.

In general, the radiative transfer obeys the following equation
\begin{equation}
    \label{eq:radtr}
    \begin{split}
        \frac{\partial}{\partial t} I_\nu (t, {\bm x}, \hat{\bm \Omega}) + \hat{\bm \Omega} \cdot {\bm \nabla}_{\bm x} I_\nu (t, {\bm x}, \hat{\bm \Omega})
        =
        & j_\nu (t, {\bm x}, \hat{\bm \Omega})
        - \lambda_{\text{abs},\nu}^{-1} \, I_\nu (t, {\bm x}, \hat{\bm \Omega}) \\
        & - \lambda_{\text{sca},\nu}^{-1}\,
        \int \dd \Omega'
        \phi_{\nu} (\hat{\bm \Omega}, \hat{\bm \Omega}')
        \qty[ I_\nu (t,{\bm x},\hat{\bm \Omega})
            -
            I_\nu ( t, {\bm x}, \hat{\bm \Omega}')
        ]\,,
    \end{split}
\end{equation}
where the source term in a direction of $\hat{\bm \Omega}$ is denoted by $j_\nu (t, {\bm x}, \hat{\bm \Omega})$,
$\lambda_{\text{abs},\nu}$ is for the absorption, and $\lambda_{\text{sca},\nu}$ is for the scatterings.
The probability of scattering is represented by $\phi_\nu (\hat{\bm \Omega}, \hat{\bm \Omega}')$ with $\phi_\nu = 1/ (4 \pi)$ for isotropic scattering.
If all the energies are stored in the relativistic plasma, the energy conservation implies that there are no absorption terms.
One may immediately confirm this by integrating Eq.~\eqref{eq:radtr} over $\Omega$ and $\nu$, which reads
\begin{equation}
    \label{eq:energy-consv}
    \frac{\partial}{\partial t} \rho (t , {\bm x})
    +
    {\bm \nabla} \cdot {\bm J} (t, {\bm x})
    = S (t, {\bm x}) - \lambda_\text{abs}^{-1}\, \rho (t, {\bm x})\,, 
    \qquad
    \lambda_\text{abs}^{-1} \equiv
    \frac{1}{\rho} \int \dd \Omega \dd \nu \,\lambda_{\text{abs}, \nu}^{-1} I_\nu\,, 
    \qquad
    S \equiv \int \dd \Omega \dd \nu\, j_\nu \,.
\end{equation}
If the source term $S (t, {\bm x})$ is driven solely by the local thermal-equilibrium plasma, we find $S (t, {\bm x}) = 0$.\footnote{
    Later we will consider the case where the high-energy radiation is present near the energetic source such as evaporating small BH.
    In this case, such a non-thermal component can be regarded as a source term.
}
Also, for the local thermal-equilibrium plasma, we should take $\lambda_\text{abs} \to \infty$ to guarantee the energy conservation.
Another important equation is obtained by integrating over $\Omega$ and $\nu$ after multiplying $\hat{\bm \Omega}$, which leaves
\begin{equation}
    \label{eq:energy-flux}
    \frac{\partial}{\partial t} {\bm J} (t, {\bm x})
    +
    {\bm \nabla} p (t, {\bm x})
    = - \lambda_\text{mfp}^{-1}\, {\bm J} (t, {\bm x})\,,
    \qquad
    \lambda_\text{mfp}^{-1}
    \equiv
    \abs{{\bm J}}^{-1} \hat{\bm J} \cdot \int \dd \Omega \dd\nu\, \lambda_{\text{sca},\nu}^{-1} \hat{\bm \Omega} I_\nu \,,
\end{equation}
with $\lambda_\text{mfp}^{-1} \sim \alpha^2 T$.
In general, $ p (t,{\bm x})$ is a tensor defined in Eq.~\eqref{eq:pressure}.
For relativistic plasma in a local thermal equilibrium, it can be simplified with the second equality in Eq.~\eqref{eq:pressure} and the equation of state implies $p = \rho/3$.
Therefore these two equations of Eqs.~\eqref{eq:energy-consv} and \eqref{eq:energy-flux} are closed and are the fundamental governing equations.

Assuming that the energy flux reaches a stationary solution much faster than the diffusion time scale, we can equate ${\bm \nabla} p (t, {\bm x})
    = - \lambda_\text{mfp}^{-1}\, {\bm J} (t, {\bm x})$.
    Substituting this into Eq. \eqref{eq:energy-consv},
we obtain the diffusion equation:
\begin{equation}
    \label{eq:diffusioneq}
    \frac{\partial}{\partial t} \rho (t , {\bm x})
    - \frac{1}{3}
    {\bm \nabla} \cdot \qty(\lambda_\text{mfp} {\bm \nabla} \rho (t, {\bm x}))
    = S (t, {\bm x}) \,.
\end{equation}
We then need to specify the source term for the system of interest, which is quite non-trivial for radiation from a BH, as we will see in Sec.~\ref{sec:thermalization}.

\subsection{Temperature profile far away from external source}
\label{subsec:Tprof}

In this section, we derive the temperature profile as a stationary solution of Eqs.~\eqref{eq:energy-consv} and \eqref{eq:energy-flux}.
As explained earlier, there are no source terms and absorption in local thermal equilibrium, and hence the governing equations are reduced to
\begin{equation}
    {\bm \nabla} \cdot {\bm J} ({\bm x}) = 0\,, \qquad
    \frac{ {\bm \nabla} \rho ({\bm x})}{3} = - \lambda_\text{mfp}^{-1} ({\bm x}) {\bm J} ({\bm x})\,.
    \label{eq:stationary}
\end{equation}
Here we make the spatial dependence of $\lambda_\text{mfp}$ explicit for clarity.

In a spherically symmetric case, the relevant quantities can be expressed as $\rho ({\bm x}) = \rho (r)$, $\lambda_\text{mfp}({\bm x}) = \lambda_\text{mfp} (r)$ and ${\bm J}({\bm x}) = J_r (r) \hat{\bm r}$ with $\hat{\bm r}$ being a unit vector along the radial direction.
The first equation in Eq.~\eqref{eq:stationary} indicates the flux conservation
\begin{equation}
    \label{eq:flux-consv}
    r^2 J_r (r) = \text{const.}
\end{equation}
The second equation implies that the energy flux is sourced by the gradients in the temperature profile.
Together with Eq.~\eqref{eq:flux-consv}, we obtain the general solution for a region without an external source such as
\begin{equation}
    \label{eq:Tprof_locth}
    \frac{\dd}{\dd r} T^3 (r) \propto r^{-2} \quad \longrightarrow \quad T (r) = \qty( \frac{r_\ast}{r} )^{1/3} \, T_\ast \,,
\end{equation}
where $T_\ast$ is determined by the boundary condition at $r = r_\ast$, \textit{i.e.,} the property of the external source placed at $r = r_s (\le r_\ast)$ as we see shortly.
We have used $\rho = ( \pi^2 g_{\ast}/30) T^4 $ where $ g_\ast $ is the effective number of degrees of freedom, which holds in a relativistic thermal plasma.

Now we can go back to
the condition of local thermal equilibrium given in Eq.~\eqref{eq:cond_locth}. Substituting Eq.~\eqref{eq:Tprof_locth} into Eq.~\eqref{eq:cond_locth}, we obtain
\begin{equation}
    \label{eq:cond_thloc_pbh}
    \alpha^2 \gg \frac{1}{T (r)\, r}
    \quad \Leftarrow \quad
    r \gg \lambda_\text{mfp} (r) \,.
\end{equation}
Therefore the local thermal equilibrium is fulfilled only for a distance away from the center by a factor of the mean free path.
This implies that
if there exists a very compact source, whose radius $r_\text{s}$ is shorter than the mean free path $\lambda_{\text{mfp}} (r_\text{s})$,
the local thermal equilibrium is not fulfilled in the vicinity of the source.

It may be instructive
to show the temperature profile for the case with a delta-function source term.
Suppose that an energy injection occurs at a certain radius $r_\ast$ and the source term is written as $S(t,{\bm x}) = S_0 \delta(r/r_\ast -1)$. 
In this case, the flux within the radius $r_\ast$ is vanishing, namely, Eq.~\eqref{eq:flux-consv} is just equal to zero.
This results in a constant temperature for $r < r_\ast$.
This solution should be continuously connected to the solution for $r>r_\ast$, which is given by Eq.~\eqref{eq:Tprof_locth}.
The temperature profile for the case with a delta-function source at $r = r_\ast$ is thus given by
\begin{equation}
 T(r) =
 \begin{cases}
  T_\ast
  \qquad &\text{for} \ \ r \le r_\ast
  \\
  T_\ast \qty(\frac{r}{r_\ast})^{-1/3}
  \qquad &\text{for} \ \ r > r_\ast
 \end{cases}\,.
\end{equation}
The overall temperature $T_\ast$ is determined by the source term such as
\begin{equation}
    \label{eq:delta_source}
    T_\ast = \qty(\frac{30}{\pi^2 g_\ast} \frac{9 \alpha^2 r_\ast^2 S_0}{4} )^{1/3} \,. 
\end{equation}

\subsection{Boundary condition from external source}
\label{sec:BC}

Now we shall consider the source term that specifies the boundary condition or $T_\ast$ in Eq.~\eqref{eq:Tprof_locth}.
Suppose that there exists an energetic spherical source with a radius being $r_\text{s}$ placed at the origin.
Depending on the radius of the source $r_\text{s}$, a qualitatively different form of the source term $S(t,r)$ should be considered.

First, let us consider the case with
$r_\text{s} \gg \lambda_\text{mfp}(r_\text{s})$.
Namely, local thermal equilibrium is established near the surface of the source.
This implies that the energy flux into the source is not negligible in addition to the energy injection into the thermal plasma from the source.
Therefore the source term $S(t,r)$ should include the incoming flux at $r = r_\text{s}$ as well as the outgoing flux from the source. 
The detail of the mathematics in this case is explained in Appendix~\ref{sec:AppA}.
The result is consistent with the intuitive result.
In particular, if the source can be regarded as blackbody radiation, the temperature $T_\ast$ at $r_\ast = r_\text{s}$ is identified as the blackbody temperature $T_\text{s}$.

Second, let us consider the case with
$r_\text{s} \ll \lambda_\text{mfp}(r_\text{s})$.
In this case,
the local thermal equilibrium is not established near the surface of the source, which implies that the energy flux into $r < r_\text{s}$ is negligible.
Then the source term $S(t,r)$ is solely determined by the energy injection from the external source.
If the energy injection from the external source per unit time is given by $\dd E_\text{s}/\dd t$,
we have
\begin{equation}
 \int^{r_\text{l-th}} \dd r' 4 \pi r'^2 S(t,r') = \frac{\dd E_\text{s}}{\dd t} \,,
 \label{eq:sourcefromE}
\end{equation}
where the local thermal equilibrium condition $r > \lambda_\text{mfp} (r)$ is saturated at $r = r_\text{l-th}$.
Here, for a moment we assume instantaneous thermalization of emitted particles at $r \simeq r_\text{l-th}$.
In particular, if the source term comes from the Stefan-Boltzmann law with temperature $T_{\rm s}$, the energy injection per unit time $\dd E_{\rm s} / \dd t$ is given by $4 \pi r_\text{s}^2 \times \pi^2 g_\ast T_{\rm s}^4 / 120$.
Note that the core temperature is much less than the blackbody temperature $T_\text{s}$ for the source.
For the stationary solution to the diffusion equation, \textit{i.e.,} $\partial \rho / \partial t = 0$, we should equate
\begin{equation}
 \int^{r_\text{l-th}} \dd r' 4 \pi r'^{2} S(t,r') = - \frac{4\pi}{3} r^2 \lambda_{\rm mfp} \frac{\partial \rho}{\partial r} \qquad \text{for}\quad r > r_\text{l-th} \,,
 \label{eq:stationarycond}
\end{equation}
which is obtained by integrating Eq.~\eqref{eq:diffusioneq}.
Combining Eqs.~\eqref{eq:sourcefromE} and \eqref{eq:stationarycond}, we obtain
\begin{equation}
    \frac{T(r)}{T_\text{s}} = \frac{3 \alpha^2}{16 \pi} \, \qty( 4 \pi r_\text{s} T_\text{s} )
    \begin{cases}
        1 &\text{for} \quad r < r_\text{l-th}\\
        \qty( r_\text{l-th} / r)^{1/3} &\text{for} \quad r > r_\text{l-th}
    \end{cases} \,,
\end{equation}
with the core radius being
\begin{equation}
    \frac{r_\text{l-th}}{r_\text{s}} = \frac{4}{3}\, \qty( \frac{\alpha^2}{4 \pi} )^{-2} \qty (4 \pi r_\text{s} T_\text{s})^{-2} \,,
\end{equation}
which is determined by $ r_\text{l-th}= \lambda_{\rm mfp}(r_\text{l-th}) = 1/(\alpha^2 T(r_\text{l-th}))$.
It is tempting to estimate the maximal temperature of the core region that could be achieved by assuming instantaneous thermalization as
\begin{equation}
    \label{eq:Tmax_l-th}
    \frac{T(r_\text{l-th})}{T_\text{s}} = \frac{3}{4} \frac{r_\text{s}}{\lambda_\text{mfp}(T_\text{s})} = \frac{3 \alpha^2}{16 \pi} \qty(4\pi r_\text{s} T_\text{s}) \,, \qquad \text{(instantaneous thermalization)}
\end{equation}
which is therefore much lower than the source temperature $T_\text{s}$ in the case of $r_\text{s} \ll \lambda_\text{mfp}$.

However, as we will discuss in the next section, the true maximal temperature is much smaller than this estimate.
This is because, for the high-energy particles with $p \sim T_\text{s} \gg T(r)$ emitted from the hot source to be thermalized locally, it takes much longer time than $\lambda_\text{mfp}$ owing to the LPM effect.
We will see in the next section how to determine the maximal temperature and the critical radius above which one may rely on $T \propto r^{-1/3}$ based on local thermal equilibrium.

\subsection{Case for a BH}
\label{sec:BH}

If the source is identified as an evaporating small BH,
we have $T_{\rm s} = T_{\rm BH}$  and $r_{\rm s} = 1/(4\pi T_{\rm BH})$.
Noting that $\lambda_{\rm mfp} \sim 1/(\alpha^2 T)$,
we obtain $r_{\rm s} \sim (\alpha^2/4\pi) \lambda_{\rm mfp}(r_{\rm s}) \ll \lambda_{\rm mfp}(r_{\rm s})$.
Interestingly, the local thermal equilibrium around the horizon of an evaporating small BH is always violated in a weakly coupled plasma $\alpha < 1$ and hence the local temperature is ill-defined there.
In this way, as long as the perturbative interactions are involved, the fundamental requirement of local thermal equilibrium immediately falsifies the argument in some studies where the temperature profile is assumed to be smoothly connected to $T_\text{s}$ at $r = r_\text{s}$.\footnote{
    A skeptical reader might further argue that a sufficiently flat profile of $T_\text{s}$ around $r \sim r_\text{s}$ would be allowed since the gradient can be small.
    However, such a flat profile immediately decays even if we put it by hand initially as it cannot be a stationary solution. See also Sec.~\ref{sec:diffusion}.
}
\footnote{
One may or may not expect that a vacuum tunneling is enhanced in the presence of a BH because of the thermal effect of Hawking radiation~\cite{Gregory:2013hja,Burda:2015isa,Burda:2015yfa,Mukaida:2017bgd,Kohri:2017ybt,Oshita:2018ptr,Shkerin:2021zbf,Shkerin:2021rhy,Strumia:2022jil}. 
See Ref.~\cite{Hamaide:2023ayu} for a recent discussion, where they used a plateau temperature profile to calculate a prefactor for tunneling rate but assumed a reduced tunneling action due to Hawking temperature.
We argue that 
the plasma temperature does not reach the source temperature $T_\text{s}$ even at $r = r_\text{s}$, and therefore, the enhancement of tunneling rate due to Hawking radiation requires further consideration. 
} 
The physical reason behind this is originated from the extreme compactness of BHs.
For a comparison, let us consider a blackbody with temperature $T_\text{s}$ whose the size is $r_\text{s}$.
If the thermal equilibrium of the blackbody is supported by their own interactions, the mean free path should be always smaller than its size, \textit{i.e.,} $r_\text{s} \gg \lambda_\text{mfp} (r_\text{s})$, which guarantees Eq.~\eqref{eq:cond_thloc_pbh} automatically.
On the contrary, BHs are so small that $r_\text{s} < \lambda_\text{mfp} (r_\text{s})$ if $\alpha < 1$.

We emphasize here that defining the local thermal temperature in general relativity is non-trivial. 
If one chooses an observer with a fixed spatial coordinate, the temperature is infinitely blue-shifted at the horizon and gravity becomes strongly coupled. 
All the discussions based on semiclassical picture of quantum field theory in curved spacetime are expected to break down in this regime.
Therefore, if we take this picture, we are forced to restrict our discussion outside the stretched horizon, slightly larger than $r_\text{s}$, where the gravity is weakly coupled.
There, our conclusion about no local-thermal equilibrium holds since gravity is weakly coupled.
To go across the stretched horizon, we need knowledge about quantum gravity.
On the other hand, if one chooses a freely-falling observer (at rest), the observed temperature, especially near the horizon, will be drastically changed, as discussed by various studies~\cite{Brynjolfsson:2008uc,Gim:2015era}. 
They argue that the temperature at the horizon becomes finite, which may indicate that the semiclassical picture can hold until the BH mass approaches the order of Planck scale. 
Assuming this argument, our low-energy effective description should be valid up to the horizon before $ T_\text{BH} $ approaches $ \Mpl $, so local thermal equilibrium cannot be achieved near the horizon. 
Nevertheless, more detailed discussion about this issue is beyond the scope of this paper, and we will focus on the regime $ r \gtrsim \order{10}r_\text{s} $ where the gravity is weakly coupled and the semiclassical treatment is justified.
We emphasize again that any perturbative dynamics cannot lead to local thermalization, and hence our conclusion holds as long as we restrict ourselves to semiclassical gravity.

In conclusion, as mentioned in the end of the previous Sec.~\ref{sec:BC}, we need to investigate the thermalization processes of high-energy particles emitted via Hawking radiation in order to estimate the temperature profile around an evaporating BH.

\section{Thermalization of Hawking radiation}
\label{sec:thermalization}

As shown in the previous section, we cannot discuss thermalization around an evaporating small BH within the assumption of local thermal equilibrium.
Based on the assumption of instantaneous thermalization for Hawking radiation, we have shown that the would-be maximal temperature given in Eq.~\eqref{eq:Tmax_l-th} is much smaller than the Hawking temperature.
However, once one takes into account the finiteness of thermalization time, the actual maximal temperature is further suppressed, as we will see shortly.
The main goal of this section is to provide a more microscopic governing equation, which describes the thermalization process of high-energy particles emitted via Hawking radiation.
In Sec.~\ref{subsec:LPM}, we first provide the qualitative estimate for the thermalization rate, taking into account the LPM effect.
In Sec.~\ref{subsec:Tprof_comb}, based on this estimate, we describe the temperature profile qualitatively by combining it with the result in the previous section.
This qualitative picture will be confirmed quantitatively in Sec.~\ref{subsec:num}.
We also review the entire picture of the evolution of the system in Appendix~\ref{app:review} for the sake of completeness.
In Sec.~\ref{subsec:kineq}, we then derive the kinetic equation including the LPM effect, which describes the microscopic processes of how the Hawking radiation deposits its energy into the ambient plasma.
The solution of the kinetic equation provides the suitable source term in Eq.~\eqref{eq:energy-consv}.
Combining the kinetic equation and Eqs.~\eqref{eq:energy-consv} and \eqref{eq:flux-consv}, we are able to investigate the thermalization around an evaporating small BH.

\subsection{In-medium cascades and the LPM effect}
\label{subsec:LPM}

Suppose that there is an energetic source with a finite size $ r_\text{s} \ll \lambda_\text{mfp}$ at the origin, which emits high-energy particles obeying a (almost) black-body spectrum with its temperature being $T_\text{s} (= T_\text{BH}$ for an evaporating BH).
Since the maximal temperature of the ambient plasma is much smaller than the source, $T \ll T_\text{s}$, the high-energy particles with its momentum being $p \sim T_\text{s}$ should reduce its momentum by cascading into lower-momentum daughters.

Owing to the hierarchy $T \ll T_\text{s}$, the daughter particles are produced almost collinearly, which requires the proper treatment of interferences among them.
The destructive interference, known as the LPM suppression~\cite{Landau:1953um,Migdal:1956tc,Gyulassy:1993hr,Arnold:2001ba,Arnold:2001ms,Arnold:2002ja,Besak:2010fb,Kurkela:2011ti}, leads to the following splitting rate:\footnote{
    See Refs.~\cite{Harigaya:2013vwa,Mukaida:2015ria,Drees:2021lbm,Drees:2022vvn,Chowdhury:2023jft} for its application to reheating after inflation.
}
\begin{equation}
    \Gamma_\text{LPM} (k,T) = c_{\rm LPM} \, \alpha^2 T \sqrt{\frac{T}{k}} \,,
    \label{eq:GammaLPM}
\end{equation}
where $ T $ is the temperature of the ambient plasma and $ k $ is the energy of the daughter particle.
The factor $ c_{\rm LPM} $ is a constant to be determined numerically later.
This time scale is larger when $ k $ is larger, so the equal splitting ($ k =p/2 $) takes the longest time while the energy loss is the largest.
One may also see that, once the primary particle of $p \sim T_\text{s}$ splits into daughters, the daughters immediately cascade down to $k \sim T$.
Therefore, we estimate the thermalization time scale as
\begin{equation}
    t_{\rm th} = \Gamma_\text{LPM}^{-1} (T_\text{s},T) \,,
    \label{eq:tth}
\end{equation}
which turns out to be much larger than $r_\text{l-th}$ in the end [see Eq.~\eqref{eq:r_lpm}].
Schematically, we expect the following evolution of distribution for emitted particles
\begin{equation}
    \label{eq:schematic}
    f (p, r \sim r_\text{s}) \simeq \frac{1}{e^{p/T_\text{s}} - 1}
    \quad
    \xlongrightarrow{\text{In-medium cascading}}
    \quad
    f (p, r \sim t_\text{th}) \simeq \frac{1}{e^{p/T} - 1} \,.
\end{equation}
Note here that we have assumed that emitted particles are massless and hence they can propagate $r = t$ with $t$ being a time after emission.

The dynamical description of this process requires the Boltzmann equation, which will be presented in Sec.~\ref{subsec:kineq}.

\subsection{Temperature profile}
\label{subsec:Tprof_comb}
Before moving on to the Boltzmann equation, here we briefly discuss the expected temperature profile combining the qualitative discussion in Sec.~\ref{subsec:LPM} and Secs.~\ref{sec:BC} and \ref{sec:BH}.

Throughout this paper, we consider the case where the time evolution of the source is much slower than the thermalization of emitted particles from the source.\footnote{
    See Appendix~\ref{app:review} for more details.
}
Hence, one may justify the stationary approximation in Sec.~\ref{sec:BC}.
As shown in Sec.~\ref{subsec:LPM}, the high-energy particles generated by the source are thermalized around $r \sim t_\text{th}$.
For qualitative discussion, let us make a crude approximation here that the high-energy particles do not thermalize at all for $r < t_\text{th}$ and do thermalize completely for $r > t_\text{th}$.
This implies that the source term is approximated by the delta function form $S (t , r) \simeq S (t, t_\text{th} (T_\text{s}, T_\ast) ) \delta (r/ t_\text{th} (T_\text{s}, T_\ast) -1) $ 
\begin{equation}
    \label{eq:source_lpm}
    \frac{\dd E_\text{s}}{\dd t} = \int \dd r'\, 4 \pi r'^2 S(t, r')
    =
     4 \pi t_\text{th}^3 (T_\text{s}, T_\ast)\, S \qty(t, t_\text{th} (T_\text{s}, T_\ast)) \,, 
\end{equation}
leading to the temperature profile given in Eq.~\eqref{eq:delta_source} with $r_\ast = t_\text{th} (T_\text{s}, T_\ast)$.
The core temperature $T_\ast$ in Eq.~\eqref{eq:delta_source} is specified by the boundary condition
\begin{equation}
    \label{eq:bdry_lpm}
     4 \pi t_\text{th}^3 (T_\text{s}, T_\ast)\, S \qty(t, t_\text{th} (T_\text{s}, T_\ast))  = -
    \frac{4 \pi}{3} r^2 \lambda_\text{mfp} \left. \frac{\partial \rho}{\partial r} \right|_{r = t_\text{th} (T_\text{s}, T_\ast) + 0^+}\,.
\end{equation}
We also assume that the source is approximated by the black-body radiation
\begin{equation}
    \label{eq:source_bb}
    \frac{\dd E_\text{s}}{\dd t} = 4 \pi r_\text{s}^2 \times \frac{\pi^2 g_\ast T_\text{s}^4}{120}\,.
\end{equation}

Combining Eqs.~\eqref{eq:source_lpm}, \eqref{eq:bdry_lpm}, and \eqref{eq:source_bb}, we obtain the following temperature profile
\begin{equation}
    \label{eq:T_lpm}
    \frac{T (r)}{T_\text{s}} \simeq c_\text{LPM}^{2/3}\, \qty(\frac{3 \alpha^2}{16 \pi})^{4/3}\, \qty( \frac{g_{\ast}}{g_{\rho \ast}} )^{2/3} \qty( 4 \pi r_\text{s} T_\text{s}  )^{4/3}
    \begin{cases}
        1 & \text{for} \quad r < r_\text{LPM} \\
        \qty( r_\text{LPM}/r )^{1/3} & \text{for} \quad r_\text{LPM} < r
    \end{cases}\,,
\end{equation}
where 
\begin{equation}
c_\text{LPM}^{2/3}\, \qty(\frac{3 \alpha^2}{16 \pi})^{4/3}\, 
\simeq 4 \times 10^{-6} 
\qty( \frac{c_{\rm LPM}}{0.4} )^{2/3}
\qty( \frac{\alpha}{0.05} )^{8/3}
 \,,
\label{eq:estimation1}
\end{equation}
with the core radius being
\begin{align}
    \label{eq:r_lpm}
    \frac{r_\text{LPM}}{r_\text{s}} = \frac{t_\text{th} (T_\text{s},T_\ast)}{r_\text{s}}
    &\simeq \frac{16}{9}\, c_\text{LPM}^{-2}\, \qty(\frac{\alpha^2}{4 \pi})^{-3}\, \qty( \frac{g_{\rho \ast}}{g_\ast} )\, \qty( 4 \pi r_\text{s} T_\text{s} )^{-3}
    \\
    &\simeq 1 \times 10^{12} 
\qty( \frac{c_{\rm LPM}}{0.4} )^{-2}
\qty( \frac{\alpha}{0.05} )^{-6}
\qty( \frac{g_{\rho \ast}}{g_{\ast}} ) 
\qty( 4 \pi r_\text{s} T_\text{s} )^{-3} \,.
\label{eq:estimation2}
\end{align}
For comparison with the literature, here we have included an additional effective degrees of freedom factor $g_{\rho \ast}$ such that $\rho = (g_{\rho\ast} \pi^2 / 30) T^4$, which can be different from $g_\ast$ in general (\textit{e.g.}, evaporating primordial black holes (PBHs).
Throughout this paper, we drop this difference for simplicity.
We will confirm this estimation by performing numerical simulations in the next Sec.~\ref{sec:simulation}.

In reality, instead of sending the initial time to the infinitely past, we have to deal with the finite time scale inherent in the system.
This temperature profile can be applicable up to a certain radius within the diffusion length
\begin{equation}
    \label{eq:rd}
    r < r_\text{d} (\tau)\,, \qquad r_\text{d} (\tau) \sim \sqrt{\lambda_\text{mfp} (r_\text{d}) \tau / 3}
    \sim \sqrt{\frac{\tau}{3 \alpha^2 T(r_\text{d})}}\,,
\end{equation}
with $\tau$ being a typical time scale.
For instance, in the case of an evaporating BH, it is identified as the evaporation time scale
\begin{equation}
    \tau = \frac{160}{\pi g_\ast} \frac{\Mpl^2}{T_\text{BH}^3}\,.
\end{equation}
For further discussion on this aspect, see Appendix~\ref{app:review}.

\subsection{Kinetic equation}
\label{subsec:kineq}

For the purpose of numerical simulation, in this subsection, we provide a detailed Boltzmann equation that governs thermalization of high-energy particles emitted from the energetic source in a background low-temperature plasma.
From the above discussion, we expect that the temperature profile is given by Eq.~\eqref{eq:T_lpm} around a BH with an almost constant mass.
The gradient of the temperature profile is then expected to be small.
In addition, we will simulate the relaxation towards the stationary solution, which is slow compared to the thermalization of high-energy particles emitted by the source.
Hence, we may include these effects in the Boltzmann equation by just replacing the temperature by $T(t,r)$.

Assuming local thermal equilibrium for a scale larger than the diffusion length, the kinetic equation for high-energy particles emitted from the source, \textit{i.e.}, Hawking radiation in the case of an evaporating BH, is written as
\begin{equation}
    \qty( \frac{\partial }{\partial r} + \frac{2}{r} ) f(p,r,t)
    = \mathcal{S}+ \left[ \mathcal{C}_{1 \leftrightarrow 2} [f] + \mathcal{C}_{2 \leftrightarrow 2} [f] \right]_{T \to T(t,r)}\,,
    \label{eq:Boltzmann0}
\end{equation}
where the time-dependence of $f(p,r,t)$ is solely induced by $T(t,r)$.
The source term $\mathcal{S}$ originates from the emission
at $r = r_\text{s}$:
\begin{equation}
    \mathcal{S} = f_\mathcal{S}(p;T_\text{s}) \delta(r-r_\text{s}) \,,
\end{equation}
where $f_\mathcal{S}$ is given by the black-body spectrum with the temperature $T_\text{s}$:
\begin{equation}
    f_\mathcal{S} (p;T_\text{s}) = \frac{1}{e^{p/T_\text{s}} - 1}\,,
    \label{eq:Hawkingspectrum}
\end{equation}
We neglect the greybody factor in the case of an evaporating BH for simplicity.
The $2\to2$ scattering term $\mathcal{C}_{2 \leftrightarrow 2} [f]$ represents elastic scatterings that are responsible for the thermalization into the background plasma for modes with small momentum $p \sim T$.
In our numerical analysis, we effectively include this effect by demanding that particles are immediately absorbed into the background plasma once their energy decrease to $p_{\rm IR} = \mathcal{O}(T)$.

The splitting term $\mathcal{C}_{1 \leftrightarrow 2} [f]$, which dominates the thermalization process for $p \gg T$, is given by~\cite{Arnold:2002zm,Mukaida:2022bbo,Mukaida:2024jiz}
\begin{equation}
    \mathcal{C}_{1 \leftrightarrow 2} [f]
    =
    \frac{(2\pi)^3}{p^2 \nu_g}  \qty[ -
        \int_0^p  \dd k \,
        \gamma_{g \leftrightarrow g g} \bigl(p; k, p-k \bigr)
        f (p)
        +
        \int_0^\infty \dd k \,
        2 \gamma_{g \leftrightarrow gg} \bigl(p+k; p, k \bigr) f (p+k)
    ]\, ,
    \label{eq:C12}
\end{equation}
under the gluon-dominance approximation for the splittings.\footnote{
    As shown in Ref.~\cite{Mukaida:2022bbo}, the in-medium cascade of high-energy SM particles asymptotically approaches a certain distribution that is independent of injected SM species and is dominated by gluon.
    For this reason, we expect that the in-medium SM splittings can be well approximated solely by gluon splittings.
    See also Ref.~\cite{Mukaida:2024jiz}.
}
On the right hand side of Eq.~\eqref{eq:C12}, $ \gamma_{g\leftrightarrow g g} (p;k,p-k) $ parametrizes the differential splitting rate of a gluon with momentum $ p $ splitting into two gluons with momenta $ k $ and $ p-k $, respectively, excluding the distribution function of the mother particle.
The explicit form of $ \gamma_{g\leftrightarrow g g} $ is given by
\begin{equation}
    \gamma_{g\leftrightarrow g g}(P; xP, (1-x)P)
    = \frac{1}{2} \frac{d_{\text{A}} C_{\text{A}} \alpha}{(2\pi)^4 \sqrt2}
    \,
    \frac{P^\text{(vac)}_{g \leftrightarrow gg} (x)}{x(1-x)}
    \,
    \mu_{\perp}^2(P; 1,x,1-x) \,, 
    \qquad
    P^\text{(vac)}_{g \leftrightarrow gg} (x) \equiv \frac{1^4+x^4+(1-x)^4}{x (1-x)}\,,
    \label{eq:gamma_ggg}
\end{equation}
and
\begin{equation}
    \mu^4_\perp \qty(P; x_1, x_2, x_3) =
    \frac{2}{\pi} \,
    x_1 x_2 x_3 \,
    P\,
    \frac{ \alpha \qty(m_\text{D}) - \alpha \qty(Q_{\perp}) }{ - b / \qty( 64 \pi^3 )}\, \mathcal{N}
    \frac{C_A}{2} \qty( x_1^2 + x_2^2 + x_3^2 )\,,
    \label{eq:mu_perp}
\end{equation}
with the dimension of adjoint representation for $\mathrm{SU}(3)$ being $d_A = 8$, its quadratic Casimir being $C_A = 3$, the degrees of freedom for gluon being $\nu_g = 16$, and the beta function of $\mathrm{SU}(3)$ being $b = -7$. 
$ P^\text{(vac)}_{g \leftrightarrow gg} (x) $ is the gluon splitting function extracting from the transverse current matrix element (see Ref.~\cite{Arnold:2002ja} for more detail).
We take $\alpha \sim 0.05$ at a typical energy (temperature) scale of our system in our numerical simulations, though its dependence comes into our final results only logarithmically.
We omit the $r$ and $t$ dependence in the distribution function and the splitting function for notational brevity.

The following quantities are written in terms of the local temperature of the background plasma $T(t,r)$:
\begin{equation}
    \mathcal{N}
    \equiv
    \sum_i
    \frac{\nu_i}{d_{i}} t_{i}
    \int \frac{\dd^3\ell}{(2\pi)^3} \,  f_\text{l-th}(\ell)
    \simeq
    \frac{15 \zeta(3) }{ \pi^2 } T^3(t,r)
    \label{eq:N_a}
\end{equation}
and
\begin{align}
    \qty( \frac{Q_{\perp}}{m_{D}} )^2
    & \sim \left(\frac{P}{T(t,r)}\right)^{1/2} \ln^{1/2}\left(\frac{P}{T(t,r)}\right) \,,
    \\
    m_\text{D}^2 & =
    8\pi \alpha
    \sum_i
    \frac{\nu_i}{d_i} t_i \int\frac{\dd^3 \ell}{(2\pi)^3} \frac{f_\text{l-th}(\ell)}{\ell}
    \simeq
    8 \pi \alpha T^2(t,r)\,,
    \label{eq:mD}
\end{align}
where $f_\text{l-th}$ is given by the local thermal distribution of the background plasma.
Note that the local-thermal plasma involves all SM contributions, while the splittings are dominated by gluon.

Using the explicit form of the splitting function, $\Gamma_{\rm LPM}(k,T)$ used in Eq.~\eqref{eq:GammaLPM} can be represented by 
\begin{equation}
\Gamma_{\rm LPM}(k,T) \simeq 
\frac{1}{\nu_g p_0} 
\gamma_{g\leftrightarrow g g}(p_0; p_0/2, p_0/2) \,.
\end{equation}
This implies 
\begin{align}
 c_{\rm LPM} \sim 0.4 \,,
\end{align}
in our setup.
We use this value as an indicator throughout this paper.

By solving the Boltzmann equation in Eq.~\eqref{eq:Boltzmann0}, we can trace the evolution of the distribution function for a given temperature profile $T(t,r)$.
As mentioned earlier, we effectively treat elastic scatterings by demanding that the soft modes with $p < p_\text{IR} \sim \mathcal{O}(T)$ are absorbed into the local-thermal plasma immediately.
This provides a source term for the diffusion equation:
\begin{equation}
    S(t,r) = \nu_g \int^{p_\text{IR}} \frac{p'^2 \dd p'}{2 \pi^2}\, p' \frac{1}{r^2} \frac{\partial}{\partial r} \qty[r^2 f(p',r,t) ]\,,
\end{equation}
for $r > r_\text{s}$.

\section{Numerical simulations}
\label{sec:simulation}

\renewcommand{\theenumi}{\Roman{enumi})}
In this section, we perform the numerical simulations and confirm Eqs.~\eqref{eq:T_lpm} and \eqref{eq:r_lpm} numerically.
In the following, we focus on the case of an evaporating BH and utilize the relation $4 \pi r_\text{s} T_\text{s}= 1$ with $r_\text{s} = r_\text{BH}$ and $T_\text{s} = T_\text{BH}$.
Summarizing the theoretical discussion in Secs.~\ref{sec:radtr} and \ref{sec:thermalization}, we have two important effects that should be taken into account:
\begin{enumerate}
\item
local thermal equilibrium is not established near the surface of the BH
\item
high-energy particles emitted as Hawking radiation can be absorbed into the ambient plasma around $r \sim t_{\rm th}$, which is 
far away from the BH
\end{enumerate}

According to the property of the diffusion equation, 
we expect that the temperature profile can be approximately expressed as
\begin{equation}
 T(r) =
 \begin{cases}
  T_\ast
  \qquad &\text{for} \ \ r \ll r_\ast
  \\
  T_\ast \qty(\frac{r}{r_\ast})^{-1/3}
  \qquad &\text{for} \ \ r \gg r_\ast
 \end{cases}\,.
 \label{eq:fit}
\end{equation}
Depending on which effect one may take into account, the parameter $T_\ast$ and $r_\ast$ vary as follows:
\begin{alignat}{3}
T_\ast &= T_{\rm BH} &\qquad r_\ast &= r_{\rm BH} &\qquad &\text{if \  I), II)
\ omitted}
\label{eq:previousT}
\\
T_\ast &= T_{\rm BH} 
\qty(\frac{9 \alpha^2}{64 \pi})^{1/3} 
\qty(\frac{r_\text{l-th}}{r_{\rm BH}})^{-1/3} &\qquad
r_\ast &= r_\text{l-th} = \frac{4  (4 \pi)^2}{3 \alpha^4}  r_{\rm BH}
&\qquad &\text{if \ II)
\ omitted}
\label{eq:ansatz3}
\\
T_\ast &= T_{\rm BH} 
\qty(\frac{9 \alpha^2}{64 \pi})^{1/3} 
\qty(\frac{r_{\rm LPM}}{r_{\rm BH}})^{-1/3} &\qquad
r_\ast &= r_{\rm LPM} = \frac{16 (4 \pi)^3}{9 c_{\rm LPM}^2 \alpha^6} r_{\rm BH} 
\label{eq:estimatedT}
\end{alignat}
where $r_{\rm LPM}$ is the thermalization length scale (or time scale) for Hawking radiation in the ambient plasma, which can be estimated as $t_{\rm th}$ given by Eq.~\eqref{eq:tth}.
The first ansatz in Eq.~\eqref{eq:previousT} is assumed in Refs.~\cite{Das:2021wei}.
The last line, Eq.~\eqref{eq:estimatedT}, is discussed in Ref.~\cite{He:2022wwy} qualitatively and will be justified numerically in Sec.~\ref{subsec:num} and \ref{sec:diffusion} of the present paper by solving the diffusion equation for the ambient plasma and Boltzmann equation for radiated particle from BH as Hawking radiation.
Even if we start from the temperature profile of different ansatz, the profile is unstable and eventually reaches the one approximated by Eq.~\eqref{eq:estimatedT}.
The latter one will be confirmed numerically as a stationary solution of the combinatory equations. 
Hence, the assumed interpolation between Eqs.~\eqref{eq:previousT} and \eqref{eq:estimatedT} in Ref.~\cite{Hamaide:2023ayu} is not stable.

\subsection{Simulations for thermalization}
\label{subsec:num}

The whole system can be described by the Boltzmann equation for high-energy particles from Hawking radiation and the diffusion equation for the locally thermalized plasma.
These are combined equations.  The Boltzmann equation depends on the background temperature, which is determined by the diffusion equation. The diffusion equation depends on the source term, which can be determined by solving the Boltzmann equation.
Nevertheless, these equations can be independently solved under a reasonable assumption as we will discuss shortly.

First, note that the temperature dependence for the splitting term is factorized such as $\mathcal{C}_{1 \leftrightarrow 2} \simeq T^{3/2}(t,r) \left[  \mathcal{C}_{1 \leftrightarrow 2} \right]_{T(t,r) \to 1}$ up to the logarithmic dependence.
Similarly, we find
$\mathcal{C}_{1 \leftrightarrow 2} \simeq T_{\rm BH}^{-1/2} \left[  \mathcal{C}_{1 \leftrightarrow 2} \right]_{T_{\rm BH} \to 1}$ by rescaling the momentum variable such as $p = T_{\rm BH} \tilde{p}$.
The Boltzmann equation we need to solve can be then written as
\begin{equation}
    \frac{\partial }{\partial \tilde{r}} F(\tilde{p},\tilde{r})
    = \mathcal{C}_{1 \leftrightarrow 2} [F]_{T \to 1}\,,
    \label{eq:Boltzmann1}
\end{equation}
where $F(\tilde{p},\tilde{r}) \equiv (r/r_\text{BH})^2 f(p,r,t)$ and $\dd\tilde{r} = T_{\rm BH}^{-1/2} T^{3/2}(t,r) \dd r$.
The boundary condition is given by $F(\tilde{p},\tilde{r}(r_\text{BH},t)) = 1 / (e^{\tilde{p}} - 1)$.
Thus an explicit $T(t,r)$ dependence can be absorbed into the redefinition of the radial coordinate.
Once we obtain the solution of $F(\tilde{p},\tilde{r})$, we can calculate $f(p,r,t)$ for any temperature profile $T(t,r)$ and $T_\text{BH}$.
This allows us to solve the Boltzmann equation without referring to the calculations for diffusion equation.
In other words, as long as there exists separation of scales, we can integrate out the short-range physics responsible for the local thermalization, and obtain the long-range physics of the fluid diffusion.

For the source term, one may again factor out the temperature profile $T(t,r)$ as follows:
\begin{equation}
    S(t,r)
    = \frac{\nu_g T_\text{BH}^{3/2} T^{3/2}(t,r)}{32 \pi^4 r^2} \frac{\tilde{S}(\tilde{p}_\text{IR},\tilde{r})}{\tilde{r}}\,,
\end{equation}
where we define
\begin{equation}
 \tilde{S}(\tilde{p},\tilde{r}) \equiv \int^{\tilde{p}} \tilde{p}'^3 \dd \tilde{p}'  \frac{\partial }{\partial \ln \tilde{r}} F(\tilde{p}',\tilde{r}) \,.
 \label{eq:tildeS}
\end{equation}
and $\tilde{p}_{\rm IR} = c_{\rm IR} T(t,r)/T_{\rm BH}$.
Therefore, once we numerically solve the Boltzmann equation \eqref{eq:Boltzmann1}, the obtained source term can be applicable to any $T(t,r)$ as long as it is slowly varying and has small gradients, \textit{i.e.,} separation of scales.

We numerically solve the Boltzmann equation.
For this purpose, we discretize
the momentum in the linear scale and
the radial coordinate in the logarithmic scale.
The relevant momentum domain in the system is $p \in (T_*, \, \mathcal{O}(T_{\rm H}))$, which implies that
we should take the resolution of momentum grid as $\Delta_{\tilde{p}} \lesssim \mathcal{O}(T_*/T_{\rm H})$.
From Eq.~\eqref{eq:tth}, the integration domain for $r$ is given by $r \in (r_{\rm BH},\, \mathcal{O}(\Gamma_\text{LPM}^{-1}(T_{\rm H})) )$.
The resolution of the radial coordinate should be smaller than the gradient of the temperature profile, such that $\Delta_{\ln \tilde{r}} \lesssim \mathcal{O}(\abs{r \partial_r \ln T(r)}) \simeq 1/3$.
According to these estimations, we take the momentum grid as $\tilde{p}_{n_p} = n_p \Delta_{\tilde{p}}$ with $\Delta_{\tilde{p}} = 5 \times 10^{-4}$, $n_p = (1,2,\dots, N_p)$, and $N_p = 2 \times 10^4$ in our numerical simulations.
Also, we take $\tilde{r}_{n_r} = \tilde{r}_0 e^{ (n_r-1) \Delta_{\ln \tilde{r}}}$ with $\tilde{r}_0 = 10^{-8}$, $n_r = (1,2, \dots, N_r)$, $\Delta_{\ln \tilde{r}} = 2 \times 10^{-3}$, and $N_r = 1.5 \times 10^4$.
Note that $e^{(N_r-1) \Delta_{\ln \tilde{r}}} \sim 10^{13}$ in this case.

\begin{figure}[t]
	\centering
 	\includegraphics[width=0.45\linewidth]{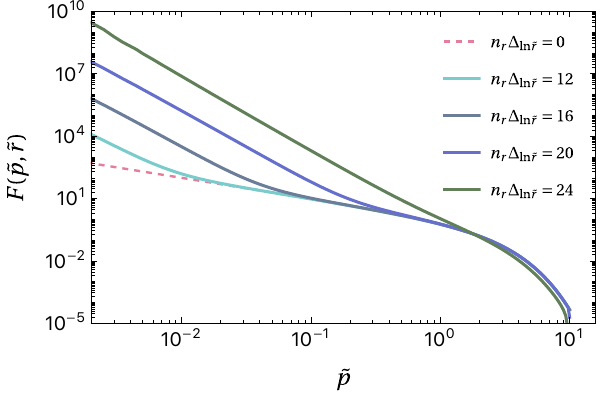}
  \quad
    \includegraphics[width=0.45\linewidth]{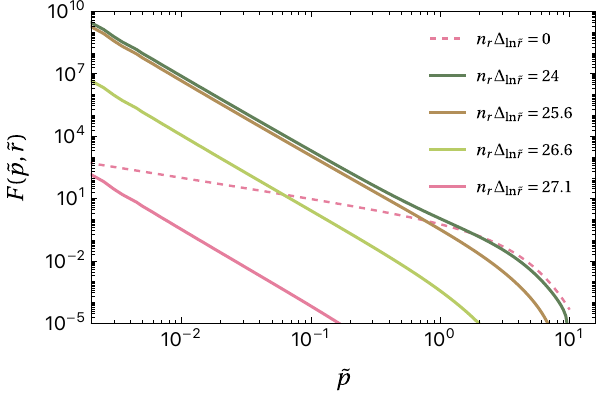}
	\caption{
        Plot for $F(\tilde{p},\tilde{r})$. The magenta dashed curve represents the initial spectrum for Hawking radiation. Note that $\tilde{r}_{n_r} = \tilde{r}_0 e^{ (n_r-1) \Delta_{\ln \tilde{r}}} \simeq \tilde{r}_0 e^{ n_r \Delta_{\ln \tilde{r}}}$ with $\tilde{r}_0 = 10^{-8}$.
    }
	\label{fig:spectrumF}
\end{figure}

Regarding $\tilde{r}$ as a time variable, we numerically solve Eq.~\eqref{eq:Boltzmann1} with "initial" distribution $F(\tilde{p},\tilde{r}(r_\text{BH},t)) = 1/(e^{\tilde{p}}-1)$ at the initial "time" $\tilde{r} = \tilde{r}(r_\text{BH},t)$.
Note again that the $t$-dependence solely stems from $T(t,r)$, which can be regarded as a constant parameter owing to the separation of scales.
As $\tilde{r}$ increases, the distribution $F(\tilde{p},\tilde{r})$ evolves.
The resulting distributions are shown in Fig.~\ref{fig:spectrumF}
for $n_r \Delta_{\ln \tilde{r}} = 12, 16, 20, 24$ (left panel) and $24, 25.6, 26.6, 27.1$ (right panel).
We show the initial distribution (\textit{i.e.}, the Bose-Einstein distribution) as the dashed curves in the figures.
The left panel shows that the distribution changes from a smaller energy scale $\tilde{p}$.
This demonstrates the bottom-up thermalization.
At a "time" around $n_r \Delta_{\ln \tilde{r}} \sim 24$ (correspondingly $\tilde{r} \sim 3 \times 10^2$), particles with energy of order $\tilde{p} = \mathcal{O}(1)$ can split into lower energy modes within a short-time scale
and therefore be thermalized.
The right panel shows that the overall distribution tends to decreases as $n_r \Delta_{\ln \tilde{r}}$ increases from $n_r \Delta_{\ln \tilde{r}} \sim 24$.
Eventually, all modes lose their energies and tend to be thermalized into the ambient plasma
for $n_r \Delta_{\ln \tilde{r}} \gtrsim 27$.
Note that $\tilde{r} \sim 3 \times 10^2$ and $5 \times 10^3$ for $n_r \Delta_{\ln \tilde{r}} =  24$ and $27$, respectively.

\begin{figure}[t]
	\centering
  \includegraphics[width=0.45\linewidth]{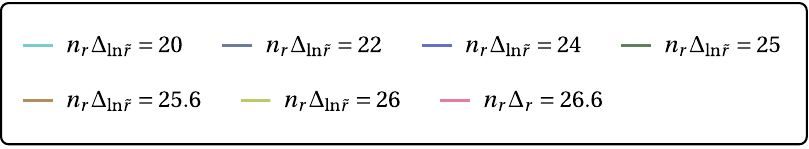}
  \\
  \vspace{0.2cm}
 	\includegraphics[width=0.45\linewidth]{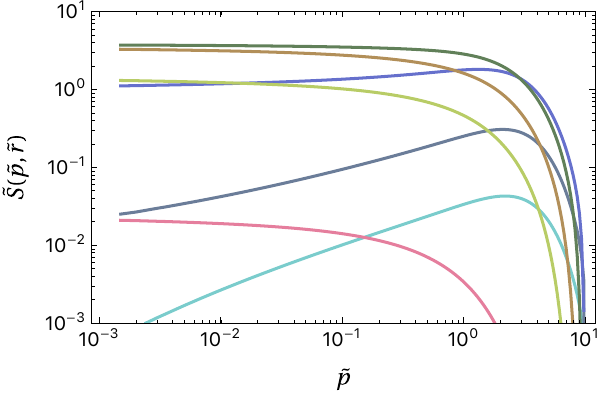}
	\caption{
        Plot for $\tilde{S}(\tilde{p},\tilde{r})$ as a function of $\tilde{p}$ for $n_r \Delta_{\ln \tilde{r}} = 20$, $22$, $24$, $25$, $25.6$, $26$, and $26.6$.
    }
	\label{fig:spectrumS}
\end{figure}

From the resulting distribution,
we can evaluate the source term from Eq.~\eqref{eq:tildeS}. For the purpose of convenience, we evaluate it for any $\tilde{p}$ and $\tilde{r}$ though the source term is given for a particular value of $\tilde{p} = \tilde{p}_\text{IR}$.
The result is shown in Fig.~\ref{fig:spectrumS} as a function of $\tilde{p}$ for $n_r \Delta_{\ln \tilde{r}} = 20$, $22$, $24$, $25$, $25.6$, $26$, and $26.6$.
One may confirm that the source term first increases up to $n_r \Delta_{\ln \tilde{r}} \sim 25$ but immediately decays after that, which is consistent with the results in Fig.~\ref{fig:spectrumF}.

\begin{figure}[t]
	\centering
 	\includegraphics[width=0.45\linewidth]{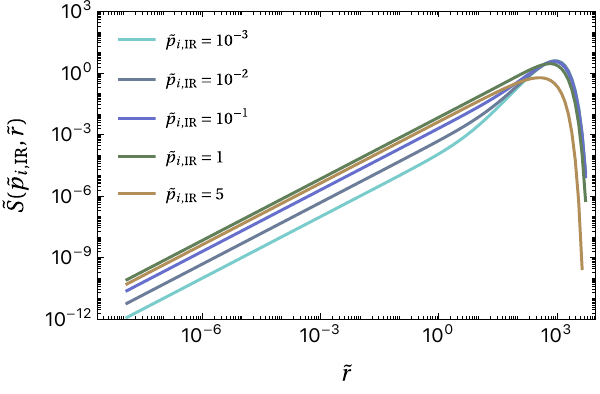}
	\caption{
        Plot for $\tilde{S}(\tilde{p}_{i,{\rm IR}},\tilde{r})$ as a function of $\tilde{r}$ for fixed $\tilde{p}_{i,{\rm IR}} = 10^{-3}$, $10^{-2}$, $10^{-1}$, $1$, and $5$.
    }
	\label{fig:spectrumS2}
\end{figure}

It may be instructive to plot $\tilde{S}(\tilde{p},\tilde{r})$ for a fixed value of $\tilde{p} = \tilde{p}_{i,{\rm IR}}$. This is shown in Fig.~\ref{fig:spectrumS2} for $\tilde{p}_{i,{\rm IR}} = 10^{-3}$, $10^{-2}$, $10^{-1}$, $1$, and $5$.
This corresponds to the energy injection in a constant temperature plasma as a function of $\tilde{r}$.
The figure shows that the energy injection is dominant for $\tilde{r} \sim 10^3$ or $n_r \Delta_{\ln \tilde{r}} \simeq 25$ for $\tilde{p}_{i,{\rm IR}} \ll \mathcal{O}(1)$. 
This is consistent with the results of Fig.~\ref{fig:spectrumF}, where the high-energy particles tend to split into lower modes and become thermalized at a "time" around $n_r \Delta_{\ln \tilde{r}} \sim 24$.
This provides a qualitative picture of the source term from the Hawking radiation.
In particular, it is almost independent of the detailed value of the IR cutoff $\tilde{p}_\text{IR}$.
We also note that
the source term is negligibly small for 
$\tilde{r} \ll 10^3$ or 
$n_r \Delta_{\ln \tilde{r}} \ll 25$.
This is what we expected: the Hawking radiation can be thermalized into the ambient plasma for a finite time scale and the source term can be relevant for a finite distance ($\sim t_\text{th} = \Gamma_{\rm LPM}^{-1})$ from the surface of the BH.

Finally, we want to emphasize that
our numerical results in this section may be useful even for other purpose because they can be applied to any temperature profile $T(t,r)$.
The information of $T(t,r)$ is absorbed into the definition of $\tilde{r}$. The actual source term can be evaluated for a given $T(t,r)$ from our results.
We particularly apply those results to the diffusion equation to understand how the Hawking radiation is thermalized around the BH in Sec.~\ref{sec:diffusion}.

\subsection{Simulations for diffusion}
\label{sec:diffusion}

Once we understand how radiation emitted as Hawking radiation is thermalized into the ambient plasma and derive the source term, we are readily solve the diffusion equation numerically.

The diffusion equation is read as Eq.~\eqref{eq:diffusioneq} or
\begin{equation}
    \frac{\partial}{\partial t} \rho (t , r)
    =
    \frac{1}{3r^2} \frac{\partial}{\partial r}  \qty[  r^2 \lambda_\text{mfp}(t, r) \frac{\partial}{\partial r} \rho (t, r)]
    + S (t, r)\,,
\end{equation}
for the spherically symmetric system.
We note that $\lambda_\text{mfp}(t,r) \sim 1/(\alpha^2 T(t,r))$ and $\rho(t,r) = \pi^2 g_\ast T^4(t,r)/30$. We take $\lambda_\text{mfp}(t,r) = 1/(\alpha^2 T(t,r))$ in our numerical simulations.
This can be rewritten as
\begin{equation}
    \frac{\partial}{\partial t} \ln T (t , r)
    =
    \frac{1}{3 \alpha^2 r^2 T} \qty[ \frac{\partial^2 \ln T}{\partial \ln r^2} + 3 \qty( \frac{\partial \ln T}{\partial \ln r})^2 +  \qty( \frac{\partial \ln T}{\partial \ln r}) ]
    + \frac{15 \nu_g T_{\rm BH}^{3/2}}{64 \pi^6 g_\ast r^2 T^{5/2}}
    \frac{\tilde{S}(\tilde{p}_\text{IR},\tilde{r}) }{ \tilde{r}}\,,
    \label{eq:diffusion1}
\end{equation}
where $\tilde{p}_\text{IR} = c_{\rm IR} T(t,r) / T_{\rm BH}$.
We take $c_{\rm IR} = 3$ in our numerical simulations.
Also, $\tilde{r}$ can be rewritten in terms of $r$ via
\begin{equation}
    \tilde{r}(r,t) = \int^r \frac{T^{3/2}(t, r')}{T_{\rm BH}^{1/2}} \dd r'\,.
    \label{eq:rtilde}
\end{equation}
The source term $\tilde{S}(\tilde{p}_{\rm IR},\tilde{r})$ is calculated by solving Boltzmann equation as shown in Fig.~\ref{fig:spectrumS}.

\begin{figure}[t]
	\centering
  \includegraphics[height=4.5cm,keepaspectratio]{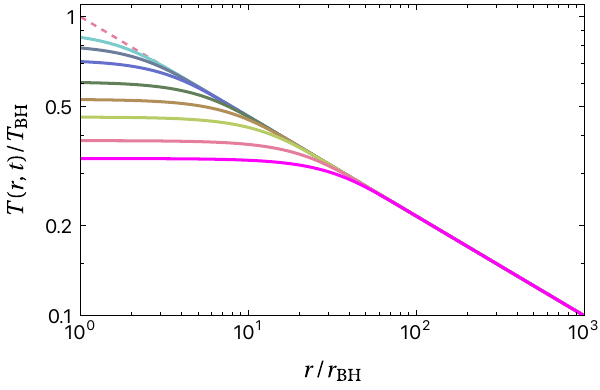}
  \quad
  \includegraphics[height=4.5cm,keepaspectratio]{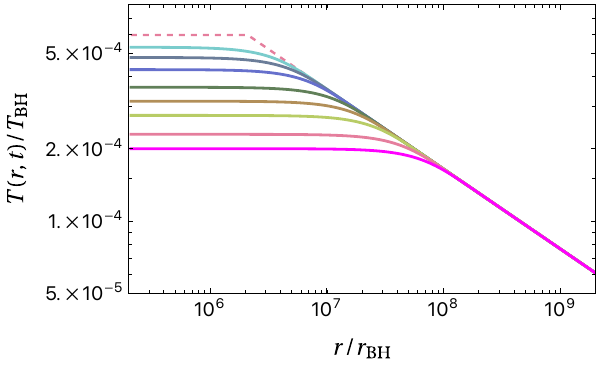}
  \\ \vspace{0.5cm}
  \includegraphics[height=4.5cm,keepaspectratio]{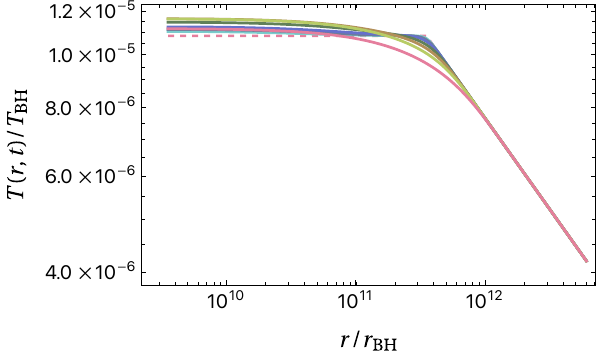}
	\caption{
        Plot for the temperature profile for different time steps.
        The initial temperature profiles, denoted as dashed lines, are given by Eqs.~\eqref{eq:previousT} (top left), \eqref{eq:ansatz3} (top right), and \eqref{eq:estimatedT} (bottom) with $r_{\rm min} = r_\ast/10$. 
        In the upper two figures, the temperature profile evolves in time from top to bottom. 
    }
	\label{fig:diffusion}
\end{figure}

We numerically solve the diffusion equation.
We discretize the radial coordinate in the logarithmic scale such that $r = r_{\rm min} e^{(n_r' - 1)\Delta_{\ln r}}$ with $\Delta_{\ln r} = 0.05$ and an integer $n_r' = 1,2,3,\dots$ ($< \mathcal{O}(100)$).
The time step for the diffusion equation should be taken such that
the left-hand side of Eq.~\eqref{eq:diffusion1} is much smaller than the first term for the right-hand side, namely,
$\Delta t \ll 3 \alpha^2 r^2 T(r) (\Delta \ln r)^2$.
This is minimal at a minimal radius $r$.
In our numerical simulation,
we take
$\Delta t = \alpha^2 r_{\rm min}^2 T(r_{\rm min}^2,t_0) (\Delta \ln r)^2 / 10$
I wonder whether this should be written as $\Delta t = \alpha^2 r_{\rm min}^2 T(t_0, r_{\rm min}) (\Delta \ln r)^2 / 10$.
The total number of time step $N_t$ is taken to be $\mathcal{O}(10^8)$.
Our algorithm to solve the diffusion equation is as follows:
\renewcommand{\theenumi}{\arabic{enumi})}
\begin{enumerate}
    \item set the initial temperature profile $T(t,r)$ at $t = 0$
    \item calculate the source term from $\tilde{S}(n_p \Delta_p, \tilde{r}_0 e^{(n_r-1) \Delta_r})$ with $n_p  \simeq c_{\rm IR} T(t,r)/(T_{\rm BH} \Delta_p)$ for every $n_r$
    \item calculate $\tilde{r}(r)$ from Eq.~\eqref{eq:rtilde} for given $T(t,r)$
    \item calculate the next time step for $T(t,r)$ from Eq.~\eqref{eq:diffusion1}
    \item repeat 2) - 4)
\end{enumerate}

It is convenient to estimate the length scale of diffusion at a given time $t$, as given in Eq.~\eqref{eq:rd}.
The diffusion length scales as $\propto \sqrt{t}$ and is as large as $(\Delta \ln r) \sqrt{N_t/30} \sim \mathcal{O}(10^2) r_{\rm min}$ for $N_t = 10^8$.
As we will see shortly, the temperature profile has a flat plateau within $r \ll r_\text{d}$
if it does not reach to the stationary solution.
We can then take $r_{\rm min}$ to be a smaller but not much smaller than the threshold of the plateau.
The explicit value will be expressed shortly for each numerical simulation.

The main purpose of the numerical simulation is to show that the temperature profile is approximately given by Eq.~\eqref{eq:estimatedT} (after considering effects I) and II) mentioned at the beginning of this section), rather than Eqs.~\eqref{eq:previousT} (without I) and II)) and \eqref{eq:ansatz3} (without II)), by demonstrating that the former one is realized as a stationary solution but the others are unstable solution to the diffusion equation. 
First, we particularly demonstrate that
the latter ansatz are unstable by taking them as initial temperature profiles. 
Figure~\ref{fig:diffusion} shows the time evolution of the temperature profile for the initial temperature profile given by Eqs.~\eqref{eq:previousT} (top left), \eqref{eq:ansatz3} (top right), and \eqref{eq:estimatedT} (bottom) with $r_{\rm min} = r_\ast/10$. 
The initial temperature profile is shown as the dashed line, which evolves into the cyan ($N_t = 5 \times 10^5$), dark green ($N_t = 10^6$), purple ($N_t = 2 \times 10^6$), green ($N_t = 5 \times 10^6$), ocher ($N_t = 10^7$), orange ($N_t = 2 \times 10^7$), pink ($N_t = 5 \times 10^7$), magenta ($N_t = 10^8$) curves, respectively. 
The first two figures demonstrate that the initial ansatz is not a stationary solution but the energy density diffuses into outer region.
Therefore the ansatz in Eqs.~\eqref{eq:previousT} and \eqref{eq:ansatz3} are not realistic temperature profiles when we consider an isolated BH and the hotspot realized after a long time since it is formed.
On the contrary, the bottom figure demonstrates that the temperature profile is relatively stable under the time evolution, which implies that the initial profile is close to the actual stationary solution to the diffusion equation. 
Still, it does not completely reach to the stationary solution that will be shown next.

\begin{figure}[t]
	\centering
  \includegraphics[width=0.45\linewidth]{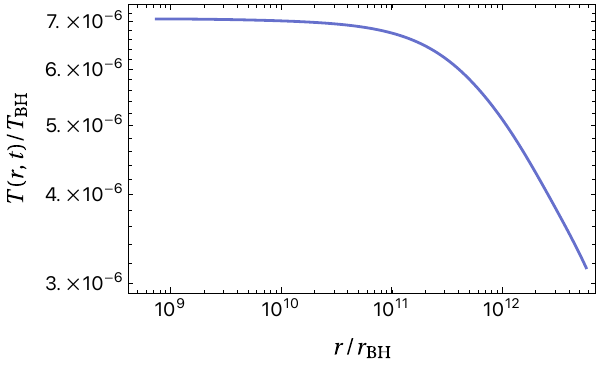}
  \quad
 	\includegraphics[width=0.45\linewidth]{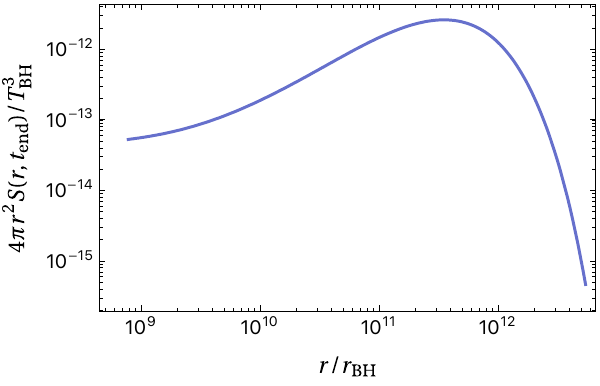}
	\caption{
        Plot for the stationary solution of the temperature profile (left) and the source term (right).
    }
	\label{fig:stationaryT}
\end{figure}

We also determine the stationary solution by evolving the diffusion equation for a long-time step and show that the resulting temperature profile is consistent with Eq.~\eqref{eq:estimatedT}.
The left panel of Fig.~\ref{fig:stationaryT}
shows the stationary solution to the diffusion equation that is realized after many time steps.
We take $r_{\rm min} = r_\ast/100$.
The resulting temperature profile has a flat plateau for $r \ll 10^{11} r_{\rm BH}$ and a power law behavior of $r \propto r^{-1/3}$ for $r \gg 10^{11} r_{\rm BH}$.
This is consistent with our qualitative discussion of Eq.~\eqref{eq:estimatedT}.
The source term for this temperature profile is shown in the right panel. As expected, it has a peak at a finite radius $r \sim 10^{11} r_{\rm BH}$ corresponding to $\Gamma_{\rm LPM}^{-1}$. This determines the regime of flat plateau for the temperature profile. 
The numerical results can be well fitted by Eq.~\eqref{eq:fit} with Eq.~\eqref{eq:estimatedT} for $T \ll T_*$ and $T \gg T_*$ within $\mathcal{O}(1)$ numerical corrections such as 
\begin{align}
 T_\ast &\simeq 7 \times 10^{-6} \, \qty( \frac{\alpha}{0.05} )^{8/3} T_{\rm BH}\,, 
\\
r_\ast &\simeq 3 \times 10^{11} \, \qty( \frac{\alpha}{0.05} )^{-6} r_{\rm BH} \,,
\end{align}
where we include the gauge coupling dependence with $\alpha = 0.05$ used in our numerical simulations. 
These results are consistent with our qualitative estimation of Eqs.~\eqref{eq:estimation1} and \eqref{eq:estimation2} with $4\pi r_s T_s= 1$ within $\mathcal{O}(1)$ numerical factors. 
These are our final results, which show that the temperature around the BH is much lower than the Hawking temperature 
and the size of the hotspot is much larger than the BH radius by many orders of magnitudes.

\section{Discussion and conclusions}
\label{sec:discussion}

In this paper, we have investigated how an evaporating BH heats up the ambient plasma quantitatively.
We clarified that the local-thermal equilibrium is never achieved around the BH radius because the extreme compactness of BH implies that the mean free path of Hawking radiation is much longer than the BH radius.
To describe the local thermalization of Hawking radiation, we first solve the Boltzmann equation numerically, and confirmed that the thermalization occurs at $t_\text{th}$, which is far from the BH radius.
Once the local-thermal equilibrium is achieved, we may switch to more macroscopic description, \textit{i.e.,} the diffusion equation of locally-thermal fluid, where the solution to the Boltzmann equation provides the source term.
We have demonstrated that the temperature profiles which do not take into account the two facts, I) local thermal equilibrium is not established around the BH radius and II) the thermalization rate of Hawking radiation is LPM suppressed, are unstable and the energy density diffuses into outer region.
On the other hand, the temperature profile based on I) and II) indeed provides the stationary solution, which are expected to be realized in a realistic situation after a long time since it is formed.
Our numerical results are consistent with our previous qualitative estimation in Ref.~\cite{He:2022wwy}.

Our analysis can be applied to any compact objects, which produce high-energy particles, such as oscillons, Q-balls, thermal-balls, gravastars, and so on. 
The qualitative feature of the temperature profile depends on the compactness of the source.
If the mean free path is much shorter than the size of the source, the temperature profile is extended up to the source surface.
Otherwise, the temperature profile has a core whose radius is greater than the source, as in the case of the evaporating BH.

The existence of locally-high-temperature regions can be a novel probe of high-energy physics, which is not expected in the standard homogeneous temperature background.
For instance, the weak Sphaleron can be reactivated locally after the electroweak phase transition, the vacuum decay can be catalyzed locally (see Refs.~\cite{Gregory:2013hja,Burda:2015isa,Burda:2015yfa,Mukaida:2017bgd,Kohri:2017ybt,Oshita:2018ptr,Strumia:2022jil} for related works), the symmetry can be restored locally, and so on.
Our discussion clarifies the regime where the local thermal equilibrium is well defined.
There we may investigate these phenomena by, for instance, utilizing the Euclidean method with its periodicity depending on the spatial radius.
It would be interesting to quantify the effects of such hotspots by taking into account the temperature gradient provided by our study.

\section*{Acknowledgement}
M.\,H.\, was supported by IBS under the project code, IBS-R018-D1.
K.\,K\, was supported by MEXT KAKENHI Grant No.\ JP23KF0289, No.\ JP24H01825, and JSPS KAKENHI Grant No.\ JP24K07027.
K.\,M.\, was supported by JSPS KAKENHI Grant No.\ JP22K14044.
M.\,Y.\ was supported by MEXT Leading Initiative for Excellent Young Researchers, and by JSPS KAKENHI Grant No.\ JP20H05851 and JP23K13092.

\appendix

\section{Temperature profile around a locally thermal equilibrated source}
\label{sec:AppA}

In this Appendix, we explain how to impose the boundary condition for a spherically symmetric source whose radius is much longer than the mean free path.
Namely, we specify the boundary condition at $r = r_\text{s}$ for $r_\text{s} \gg \lambda_\text{mfp}$.
This is not the case for a BH, but we explain this for the sake of consistency and completeness.

Suppose that the energetic source term is a blackbody of temperature $T_\text{H}$.
Since local thermal equilibrium is maintained everywhere in this case, we should be able to use Eq.~\eqref{eq:stationary} even at $r = r_\text{s}$.
Integrating the second equation of Eq.~\eqref{eq:stationary} from $r_\text{s}$ to $r_\text{s} + \epsilon$ with $0 <\epsilon /r_s \ll 1$, we obtain
\begin{equation}
    \frac{\rho (r_\text{s} + \epsilon) - \rho (r_\text{s})}{3 \epsilon} = - \lambda_\text{mfp}^{-1} (r_\text{s})J_r (r_\text{s}) + \mathcal{O} ( \epsilon ), \qquad
    \rho (r_\text{s}) = \frac{\pi^2 g_\ast}{30} T_\text{H}^4\,,
\end{equation}
where $g_\ast$ being the effective relativistic degrees of freedom.
To have a finite flux, the left-hand side should have a well-defined limit, which implies that the temperature should be continuous at $r = r_\text{s}$:
\begin{equation}
    \rho (r_\text{s} + 0^+) = \frac{\pi^2 g_\ast}{30} T_\text{H}^4 \quad \Rightarrow \quad
    T(r_\text{s}) = T_\text{s} = T_\text{H} \,.
\end{equation}
The energy flux at $r_\text{s}$ is given by the right derivative of the energy density
\begin{equation}
    \label{eq:app_bb-flux}
    J_r (r_\text{s})
    = -\frac{\lambda_\text{mfp} (r_\text{s})}{3} \rho' (r_\text{s} + 0^+)
    = \frac{16}{9} \frac{\lambda_\text{mfp} (r_\text{s})}{r_\text{s}}
    \times J_{\text{SB},r}\,,
    \qquad 
    J_{\text{SB},r} \equiv \frac{\pi^2 g_\ast}{120} T_\text{H}^4 \,,
\end{equation}
where we have used Eq.~\eqref{eq:Tprof_locth}.
The energy flux is much smaller than that predicted by the Stefan--Boltzmann law with temperature $T_\text{H}$, $J_{\text{SB},r}$, for $\lambda_\text{mfp} \ll r_\text{s}$.
This is simply because there exists a significant incoming flux into the blackbody from the ambient plasma, whose contribution almost cancels out the outgoing $J_{\text{SB},r}$, leaving a suppressed outgoing flux of $(\lambda_\text{mfp}/r_\text{s}) J_{\text{SB},r}$.

However, one should note that this is not the case for a BH because its radius is much shorter than the mean free path in the ambient plasma as we discuss in the main part of this paper.

We note that, in a realistic situation, the temperature of the source would evolve in time owing to the energy conservation.
Nevertheless, as long as the time-dependence of the source is much slower than the relaxation time $\sim 1/(\alpha^2 T)$, this stationary solution can approximate the true evolution at least within $r < [\tau_\text{s}\, \lambda_\text{mfp}(r)]^{1/2}$ with $\tau_\text{s}$ being a typical timescale of the source-term evolution.

\section{Review of qualitative estimation in previous study}
\label{app:review}
Here we review the entire picture of the evolution in our previous paper~\cite{He:2022wwy} for the sake of completeness.
In~\cite{He:2022wwy}, we have taken into account the evolution of the source, and discuss its effect on the temperature profile qualitatively.

Let us first start with the intrinsic time scale of the source evolution.
Due to Hawking radiation, the mass loss rate of a black hole with mass $ M $ is given by
\begin{align}
    \frac{\dd M}{\dd t} = -\frac{\pi}{480} g_{H*}(T_{\rm H}) \frac{\Mpl^4}{M^2} \,,
\end{align}
where $ g_{H*} $ is the effective number of degrees of freedom of the Hawking radiation and we have neglected the greybody factor for simplicity.
This gives the evaporation time scale of the black hole
\begin{equation}\label{eq-tevap}
    t_{\rm evap} (M) = \frac{160}{\pi g_{H*}}\frac{M^3}{\Mpl^4} = \frac{160}{\pi g_{H*}}\frac{\Mpl^2}{T_{\rm BH}^3}\,,
\end{equation}
which characterizes the $\mathcal{O}(1)$ depletion timescale of the evaporating BH.
Here we have used $T_\text{BH} = \Mpl^2 / M$.

As discussed in Sec.~\ref{subsec:Tprof_comb}, the time evolution of the source sets an absolute upper bound on the length scale within which the ambient thermal plasma can keep up with the evolution of the source.
Inserting the evaporation time scale to Eq.~\eqref{eq:rd}, we obtain the decoupling radius:
\begin{equation}
    \frac{r_\text{dec} (M)}{r_\text{BH}(M)} \equiv \frac{r_\text{d} (t_\text{evap} (M))}{r_\text{BH}(M)}
    \simeq 13 \, \alpha^{-8/5}\qty(\frac{g_{\rho\ast}}{106.75})^{-3/5} \qty( \frac{g_{\rho\ast}}{g_\ast} )^{4/5} \qty( \frac{M}{\Mpl} )^{6/5} \,.
\end{equation}
This is much larger than the initial core radius $r_\text{LPM}$ given in Eq.~\eqref{eq:r_lpm} for a sufficiently heavy BH with a perturbative coupling, \textit{e.g.,} $M \gtrsim 0.4 \,\mathrm{g}$ and $\alpha \simeq 0.1$.

As the BH mass starts to decrease by $\mathcal{O}(1)$ via the Hawking radiation for $t \gtrsim t_\text{evap}(M)$, $r_\text{dec}/r_\text{BH}$ decreases while $r_\text{LPM} / r_\text{BH}$ remains constant.
The local thermal plasma for $r_\text{dec} (M) \lesssim r \lesssim r_\text{dec} (M_\text{ini})$ is decoupled from the BH evolution and its temperature freezes-out.
The temperature profile at this stage can be expressed as
\begin{equation}
    \frac{T (r)}{T_\text{BH} (M)} \simeq c_\text{LPM}^{2/3}\, \qty(\frac{3 \alpha^2}{16 \pi})^{4/3}\, \qty( \frac{g_{\ast}}{g_{\rho \ast}} )^{2/3}
    \begin{cases}
        1 & \text{for} \quad r < r_\text{LPM} (M)\,,\\[.5em]
        \qty(\frac{r_\text{LPM}(M)}{r}  )^{1/3} & \text{for} \quad r_\text{LPM} (M) < r < r_\text{dec} (M) \,, \\[.5em]
        \qty( \frac{r_\text{LPM}(M)}{r_\text{dec}(M)})^{1/3}  \qty( \frac{r_\text{dec}(M)}{r})^{7/11} & \text{for} \quad r_\text{dec} (M) < r <r_\text{dec} (M_\text{ini}) \,.
    \end{cases}
\end{equation}
We note that the prefactor can be estimated as 
\begin{equation}
c_\text{LPM}^{2/3}\, \qty(\frac{3 \alpha^2}{16 \pi})^{4/3}\, 
\qty( \frac{g_{\ast}}{g_{\rho \ast}} )^{2/3}
\simeq 4 \times 10^{-6} 
\qty( \frac{c_{\rm LPM}}{0.4} )^{2/3}
\qty( \frac{\alpha}{0.05} )^{8/3}
\qty( \frac{g_{\ast}}{g_{\rho \ast}} )^{2/3} \,.
\end{equation}
The thermalization length scale $r_{\rm LPM}$ can be estimated as 
\begin{equation}
    \frac{r_\text{LPM}}{r_\text{BH}} 
    \simeq \frac{16}{9}\, c_\text{LPM}^{-2}\, \qty(\frac{\alpha^2}{4 \pi})^{-3}\, \qty( \frac{g_{\rho \ast}}{g_\ast} ) 
    \simeq 
    1 \times 10^{12} 
\qty( \frac{c_{\rm LPM}}{0.4} )^{-2}
\qty( \frac{\alpha}{0.05} )^{-6}
\qty( \frac{g_{\rho \ast}}{g_{\ast}} ) \,.
\end{equation}

After the evaporation proceeds sufficiently, $r_\text{dec}$ becomes smaller than $r_\text{LPM}$, which occurs at
\begin{equation}
    M \lesssim M_\ast \simeq 10^2 \, c_\text{LPM}^{5/3} \, \alpha^{-11/3}\, \qty( \frac{g_{\rho\ast}}{106.75} )^{1/2} \qty( \frac{g_{\rho\ast}}{g_\ast} )^{1/6} \Mpl \,.
\end{equation}
The high-energy particles emitted for $M \lesssim M_\ast$ does not thermalize and penetrate through the hotspot, and hence the temperature profile freezes-out at $M \simeq M_\ast$.
The temperature profile for $M \lesssim M_\ast$ thus reads
\begin{equation}
    T(r) \simeq T_\text{max} 
    \begin{cases}
        1 &\text{for} \quad r < r_\text{LPM} (M_\ast)\,, \\
        \qty( \frac{r_\text{LPM} (M_\ast)}{r} )^{7/11} &\text{for} \quad r_\text{LPM} (M_\ast) < r < r_\text{dec} (M_\text{ini}) \,,
    \end{cases}
\end{equation}
with the maximal temperature being
\begin{align}
    T_\text{max} &= 2 \times 10^{-4}\, c_{\text{LPM}}^{7/3} \, \alpha^{19/3} \, \qty(\frac{g_{\rho\ast}}{106.75})^{-1/2} \qty(\frac{g_\ast}{g_{\rho\ast}})^{5/6} \Mpl \\
    &\simeq 3 \times 10^{5}\,\mathrm{GeV}\,\qty(\frac{c_{\text{LPM}}}{0.4})^{7/3} \, \qty(\frac{\alpha}{0.05})^{19/3} \, \qty(\frac{g_{\rho\ast}}{106.75})^{-1/2} \qty(\frac{g_\ast}{g_{\rho\ast}})^{5/6}\,.
\end{align}

Finally, after the BH completely evaporates, the source term in the diffusion equation disappears, and then the hotspot diffuses into the bulk thermal plasma.

This is the qualitative estimate of the hotspot around an evaporating black hole done in our previous work~\cite{He:2022wwy}.


\small
\bibliographystyle{utphys}
\bibliography{ref}

\end{document}